# Field-Free Switching in Symmetry Breaking Multilayers – The Critical Role of Interlayer Chiral Exchange

Yung-Cheng Li[1†], Yu-Hao Huang[1†], Chao-Chung Huang[1], Yan-Ting Liu[1], and Chi-Feng Pai[1,2,3*]

[1]*Department of Materials Science and Engineering, National Taiwan University, Taipei 10617, Taiwan*

[2]*Center of Atomic Initiative for New Materials, National Taiwan University, Taipei 10617, Taiwan*

[3]*Center for Quantum Science and Engineering, National Taiwan University, Taipei 10617, Taiwan*

It is crucial to realize field-free, deterministic, current-induced switching in spin-orbit torque magnetic random-access memory (SOT-MRAM) with perpendicular magnetic anisotropy (PMA). A tentative solution has emerged recently, which employs the interlayer chiral exchange coupling or the interlayer Dzyaloshinskii-Moriya interaction (i-DMI) to achieve symmetry breaking. We hereby investigate the interlayer DMI in a Pt/Co multilayer system with orthogonally magnetized layers, using repeatedly stacked [Pt/Co]$_n$ structure with PMA, and a thick Co layer with in-plane magnetic anisotropy (IMA). We clarify the origin and the direction of such symmetry breaking with relation to the i-DMI's effective field, and show a decreasing trend of the said effective field magnitude to the stacking number (n). By comparing the current-induced field-free switching behavior for both PMA and IMA layers, we confirm the dominating role of i-DMI in such field-free switching, excluding other possible mechanisms such as tilted-anisotropy and unconventional spin currents that may have arisen from the symmetry breaking.

---

[†] Y.-C. L. and Y.-H. H. contributed equally to this work
[*] Email: cfpai@ntu.edu.tw

# I. Introduction

Making industrially viable spin-orbit torque (SOT) based magnetic random-access memory (MRAM) has been at the forefront of spintronics research for the past decade. For this goal, great endeavors have focused on improving various key parameters such as the efficiency of spin generation [1-3], readability [4-6] and thermal stability [7-9], …etc. The most critical issue, however, is the conundrum that magnetizations with perpendicular magnetic anisotropy (PMA) cannot be deterministically controlled by a conventional spin current with in-plane polarization due to the symmetry constraint, unless some sort of symmetry breaking element is introduced, such as an in-plane field parallel to the applied current [10,11]. This issue significantly hampers the feasibility for building SOT-MRAMs with PMA, which is essential in developing high density magnetic memories [7,12]. Numerous methods have been proposed to make the switching deterministic and integrable with conventional magnetic tunnel junctions with PMA (p-MTJs), such as anisotropy engineering through wedged structures [13-17], unconventional spin current generations [18,19], exchange biased structures [20,21], and the magnetic hard mask approach [22]. These proposals often generate new shortcomings, however. For example, dipolar coupling/exchange coupling's switching polarity may be magnetic history dependent. Unconventional spins require more exotic fabrication procedures and are sometimes magnetic history correlated as well [19,23].

One alternative approach has been proven to be feasible lately, which involves harnessing the newly discovered interlayer chiral exchange or interlayer Dzyaloshinskii-Moriya interaction (i-DMI). DMI is an antisymmetric exchange interaction which favors perpendicular arrangement of adjacent spins [24-27]. Contrary to the more conventional interfacial case of DMI, the interlayer version mediates orthogonal spin configurations between separate magnetic layers, rather than within the same magnetic layer. Recently, after being predicted by Monte Carlo calculations [28] and experimental verifications of i-DMI in double PMA and synthetic antiferromagnetic (SAF) systems [29,30], several seminal works had further utilized it to achieve current-induced field-free magnetization switching [31,32]. The advantages of exploiting i-DMI to realize field-free switching within PMA systems are manifold. Since i-DMI has been proven to exist in both orthogonally magnetized and SAF systems, the versatility of stack engineering may be significantly increased when compared to the conventional interfacial case. The magnetization switching behavior is also governed by the characteristic DMI vector **D**, which defines both the strength and direction of the chiral exchange, and subsequently the switching polarity. Harnessing i-DMI to achieve field-free switching also eliminates the possibility of bipolar switching since the magnetization configuration reinitializes itself under the applied current [33]. These features are all essential for realizing practical SOT-MRAM.

The Pt/Co multilayer system has already been shown to be important for magnetic memory applications [34,35]. In this work, we systematically study the feasibility to incorporate i-DMI into heterostructures with multilayers having in-plane magnetized Co and perpendicularly magnetized [Pt/Co]$_n$ and reveal the correlation between the obliquely-grown [Pt/Co]$_n$ and the i-DMI strength. The i-DMI-induced effective field is found to decrease while increasing the stacking number n of the [Pt/Co]$_n$ structure. Furthermore, a non-negligible tilted anisotropy is also observed in such symmetry-broken multilayer system. We also demonstrate current-induced field-free magnetization switching and show that the switching percentage drops as the i-DMI weakens, which confirms the dominating role of i-DMI over the tilted anisotropy in the observed field-free switching. Current-induced loop shift measurement results, finally, further exclude other mechanisms that might lead to field-free switching, such as unconventional z-polarized spin current, and reconfirm the critical role of i-DMI in Pt/Co multilayer system.

## II. Sample preparation and characterization

The samples used in this work are prepared by magnetron sputtering under a base pressure of ~$10^{-8}$ Torr, with structures being Ta(0.5)/[Pt(1)/Co(0.8)]$_n$/Pt(2.2)/Co(1.7)/Ta(3) deposited on thermally oxidized Si wafers (n = 1, 2, ,3, 4, and the unit of the numbers in parentheses is nm). Ta(0.5) and Ta(3) serve as the adhesion and the capping layer, respectively. As schematically shown in Fig. 1(a), [Pt/Co]$_n$ multilayers have PMA while the top Co(1.7) layers in all samples have in-plane magnetic anisotropy (IMA). Since our magnetic multilayers are presumably polycrystalline, high symmetry is expected for all layers under normal sputtering conditions. However, throughout deposition, the substrate's rotation is disabled until the Pt(2.2) spacer which serves as the chiral exchange coupling media located between [Pt/Co]$_n$ multilayers and Co in-plane layer was grown (For discussions on the choice of the Pt spacer thickness to be 2.2 nm, see Appendix A). Such oblique deposition technique is used to induce symmetry breaking, whose direction is mainly dominated by the atom flow direction of underlayer Ta of 25°, as seen Fig. 1(a), due to the templating effect it could bring about which is further attributed to the formation of canted columnar structure by oblique deposition of the seed layer [15,36]. To achieve current-induced field-free switching, we make the symmetry breaking direction along *y* axis, that is, transverse to the applied current (*x*) direction [13-17,32]. With the only remaining symmetry operation being a mirror along the oblique-deposition direction, i-DMI's effective field should be symmetrical across this mirror, thus, the DMI **D** vector has to point toward a direction perpendicular to the wedge direction [33,37], just as we will demonstrate in the next section.

For electrical measurements, micron-sized Hall bar devices are patterned through lift-off process with lateral dimensions of 5μm by 60μm. We measure anomalous Hall resistance $R_H$ and unidirectional magnetoresistance (UMR) from these Hall bar devices to observe PMA and IMA magnetization states with direct sensing current $I_{DC}$, respectively [38,39]. The i-DMI characterization is carried out by sweeping out-of-plane magnetic field $H_z$ under static in-plane magnetic field $H_{IP}$ along different in-plane angle $\varphi_H$ via a well calibrated vector electromagnet which creates accurate magnetic fields, as Fig. 1(b) depicts (please refer to Ref. [40] for the details of the calibration technique). Representative $R_H$ hysteresis loop of a [Pt/Co]$_n$ multilayer sample with n = 1 is shown in Fig. 1(c). In comparison, shifts of the hysteresis loops due to the i-DMI effective field is observed when $H_{IP}$ is introduced, as shown Fig. 1(d). Also note that the coercivity $H_C$ of the [Pt/Co]$_n$ multilayer increases with increasing the stacking number n (listed in Table I) due to the enhancement of interfacial anisotropy generated by improved fcc texture [41,42].

## III. i-DMI coupling between [Pt/Co]$_n$ and Co

To quantify the interlayer chiral exchange coupling induced by orthogonal ferromagnetic configuration with symmetry breaking, $H_{IP}$ is applied along different $\varphi_H$ to magnetize the IMA Co layer, thereby inducing an i-DMI effective field acting upon the PMA [Pt/Co]$_n$ layer along $z$ axis, $H_z^{shift}$, which can be estimated through the offset in the out-of-plane hysteresis loops. Subsequently, Fig. 2(a) demonstrates that such $H_z^{shift}$ for the wedged PMA [Pt/Co]$_{n=1}$ layer exhibits a sinusoidal variation with regard to $\varphi_H$ under $H_{IP}$ = 100 Oe and the maximum (minimum) values of $H_z^{shift}$ are approximately located at $\varphi_H$ = 90° (270°), which corresponds to the direction of the oblique growth and is perpendicular to the vector **D** (**D** along $x$, $\approx$ 0°) in this broken symmetry system. The rest of the main samples possess identical angle dependence of $H_z^{shift}$. This sinusoidal dependence is in stark contrast to the control sample (structurally similar to n = 1 but grown with the sample holder rotation on) results as also shown in Fig. 2(a), a minimal $H_z^{shift}$ at all $\varphi_H$ angles is demonstrated, showcasing the lack of i-DMI.

This antisymmetric variation is attributed to the i-DMI effect instead of typical dipolar field or symmetrical interlayer exchange coupling [43-45]. In this system, exchange energy term $E_{ex}$ can be expressed as $E_{ex}$ = -$J_H$**M**$_1$·**M**$_2$ - **D**·**M**$_1$×**M**$_2$ [26,46]. The first term represents conventional Heisenberg symmetric exchange term. The second term describes the antisymmetric exchange term, *i.e.*, the i-DMI contribution. $J_H$ and **M**$_{1,2}$ represent the conventional Heisenberg exchange constant and the magnetization of the two magnetic layers, respectively. When a magnetic field is applied, Zeeman energy $E_{Zeeman}$ = - **M**·**H**$_{ext}$ is also taken into consideration. Therefore, in our structure, if $M_1$ and $M_2$ are assigned to be $M_{PMA}$ ([Pt/Co]$_n$) and $M_{IMA}$ (Co), the overall field acting upon the PMA layer can be expressed as $H_z^{eff} = H_{ext} - D \times M_{IMA}$ in the absence of $J_H$. $H_z^{shift}$ then corresponds to the magnitude of $-D \times M_{IMA}$.

Next, by performing in-plane field $H_{IP}$ scans with its direction fixed at $\varphi_H = 90°$, we observe that $H_z^{shift}$ vs. $H_{IP}$ (or $H_y$) can be divided into two regions with different slopes, denoted as the shaded and the white sections in Fig. 2(b). The slope transition points $H_{IP}^{trans}$ are approximately located at $H_{IP} = 75$ Oe for sample with n = 1 and 100 Oe for n = 2, 3, and 4, indicating that i-DMI's contribution to $H_z^{shift}$ reaches saturation. These transition points also correspond to $M_{IMA}$ being fully aligned toward $\varphi_H = 90°$ under sufficient $H_{IP}$. Beyond $H_{IP}^{trans}$, $H_z^{shift}$ still increases linearly with increasing $H_{IP}$. This additional $H_z^{shift}$ is attributed to the titled anisotropy due to the wedged structure [16,17], which will be further examined in the next section. Note that the control sample shows a minimal $H_z^{shift}$ in Fig. 2(b) and its absence of a slope change under different $H_{IP}$ again demonstrates the lack of both i-DMI and tilted anisotropy without symmetry breaking.

By considering the measured $H_z^{shift}$ to be the superposition of both i-DMI and tilted anisotropy contributions, and the fact that the tilted anisotropy's contribution as a function of $H_{IP}$ is linear [16,47], maximum value of the effective i-DMI field $H_{z,DMI}^{sat}$ can be extracted from subtracting the $H_z^{shift}$ at $H_{IP}^{trans}$ ($H_z^{shift, sat}$) by the products of corresponding $H_{IP}^{trans}$ and $H_z^{shift}/H_y$ at white regions. It is found that the i-DMI $H_{z,DMI}^{sat}$ gets attenuated when increasing the stack number n, as shown in Fig. 2(c). Note that the value of $H_{z,DMI}^{sat}$ for sample n = 1 is close to the result in a previous study with a similar Pt interlayer thickness [48]. More details regarding the Pt spacer thickness dependence of i-DMI can be found in Appendix A. As further shown in Fig. 2(d), the $H_z^{\text{shift}}/H_y$ governed by i-DMI coupling (slopes from the shaded region in Fig. 2(b)) consequently possesses a decaying trend with n, which is due to the limited interaction length from $\mathbf{M}_{\text{IMA}}$ with Pt(2.2) as the mediating layer. On the other hand, the values of $H_z^{\text{shift}}/H_y$ governed by the tilted anisotropy (slopes from the white region in Fig. 2(b)) show a less obvious but opposite trend, increasing from 0.04 to 0.085 for sample n =1 to 4. Compared to a previous work on strain-induced tilted anisotropy, the variation of $H_z^{\text{shift}}$ to $H_{\text{IP}}$ in Fig. 2(b) shows a distinct difference, while the values of tilted anisotropy induced $H_z^{\text{shift}}/H_y$ are of the same scale [49].

### IV. Estimation of tilted anisotropy

Despite a smaller effect when compared to the i-DMI, tilted magnetic anisotropy's contribution to hysteresis loop shift is non-negligible and may potentially play a role in assisting current-induced field-free switching. Thus, to further quantify the tilted anisotropy, $H_z^{\text{shift}}/H_{\text{IP}}$ in large field regime (white region) is measured under different static $H_{\text{IP}}$ with varying $\varphi_H$, as shown in Fig. 3(a) for the n = 2 sample grown obliquely. Compared to the n = 2 sample without wedge, *i.e.*, the one deposited with rotation throughout the entire sputter process, it is clear that there exists a significant difference of $H_z^{\text{offset}}/H_{\text{IP}}$ variation with $\varphi_H$ measured at large field, which is summarized in Fig. 3(b). These $H_z^{\text{offset}}/H_{\text{IP}}$ vs. $\varphi_H$ data can be further fitted to extract the easy axis tilted angle $\theta_{\text{ani}}$ (away from the z-axis) by [16,17],

$$H_z^{\text{offset}}\cos\theta_{\text{ani}}=H_{\text{IP}}\cos(\varphi_{\text{ani}}-\varphi)\sin\theta_{\text{ani}}, \tag{1}$$

where $\theta_{\text{ani}}$ in our main sample with n = 2 (wedged) is approximately 4.4° whereas only 0.4° in the control sample with n = 2 (non-wedged). Furthermore, $\theta_{\text{ani}}$ slightly increases with n, as shown in Fig. 3(c). This is tentatively attributed to the template effect provided by the Ta buffer layer [36] being strengthened as the number of stacks of [Pt/Co]$_n$ without rotation increased. Note that $\varphi_{\text{ani}}$ indicates the in-plane angle that **M**$_{\text{PMA}}$ tilts toward. As Fig. 3(b) exhibits, the maximum of $H_z^{\text{offset}}/H_{\text{IP}}$ for sample n = 2 is located at $\varphi_H = 90°$, showing that $\varphi_{\text{ani}}$ is close to 90°, corresponding to the wedge direction. Additionally, the magnitudes of the wedge-induced $\theta_{\text{ani}}$ found in this work are comparable to other studies' values obtained from similar wedged structures with tilted anisotropy (3.3° in [16] and 2.6° in [47]).

Furthermore, with the occurrence of a tilted **M**$_{\text{PMA}}$, the Heisenberg exchange contribution to $H_z^{\text{shift}}$ may need further scrutiny. Based on typical values of $M_s^{\text{Co}} = 1.18 \times 10^6$ A/m [50], $M_s^{\text{Pt-Co multilayers}} = 1.8 \times 10^6$ A/m [51] and the Heisenberg interlayer exchange energy areal density of $E_{\text{ex}} \cong 2 \sim 5$ μJ/m$^2$ reported in similar structures and Pt interlayer thicknesses [52-54], these values are plugged into $E_{\text{ex}} = \mu_0 M_s H_{\text{ex}} t_{\text{FM}}$ to obtain the net effective field $H_{\text{ex}}$. Even when using the highest value of $\theta_{\text{ani}}$ in our main samples, which is 5.6° for n = 4, the net effective field $H_{\text{ex}}$ applied along canted **M**$_{\text{PMA}}$ would still be less than 4 Oe. In sharp contrast, when applying previously reported i-DMI energy areal density of $E_{\text{DMI}} \cong 24 \sim 44$ μJ/m$^2$ [33,48] into $E_{\text{DMI}} = \mu_0 M_s H_{\text{DMI}} t_{\text{FM}}$, the calculated $H_{\text{DMI}}$ ranged from 40 to 80 Oe, which are in much better agreement with xour observed magnitude of the i-DMI effective fields. Therefore, the anti-symmetric i-DMI mechanism still dominates over the symmetric Heisenberg exchange even in the presence of tilted anisotropy.

Another evidence of the minimal contribution from the symmetric Heisenberg exchange is the fact that when n increases from 1 to 4, the measured $H_{z,\text{DMI}}^{\text{sat}}$ decreases from 101 to 34 Oe, while the measured $\theta_{\text{ani}}$ increased from 2.4 to 5.6°. This significant $H_{z,\text{DMI}}^{\text{sat}}$ decrease while any contributions from Heisenberg exchange should have increased more than twofold (due to the increase in $\theta_{\text{ani}}$) suggests that the Heisenberg exchange effective field plays a minor role in contributing to $H_{z,\text{DMI}}^{\text{sat}}$.

## V. Current-induced field-free switching with i-DMI

Next, we examine the feasibility of employing i-DMI for current-induced field-free switching of $\mathbf{M}_{\text{PMA}}$, by applying pulsed current $I_{\text{write}}$ with pulse width of 50 ms along longitudinal direction of the Hall bar devices and detect the switching of $\mathbf{M}_{\text{PMA}}$ by means of $R_H$. On the other hand, switching dynamics of $\mathbf{M}_{\text{IMA}}$ is simultaneously probed by the UMR, which is recorded through longitudinal resistance difference $\Delta R$ as sensed by $I_{\text{DC}}$ with opposite polarities [55,56]. From the symmetry perspective, the writing current is applied parallel to the **D**-vector $I_{\text{write}} \parallel \mathbf{D}$ ($I_{\text{write}} \perp \mathbf{M}_{\text{IMA}}$) such that the symmetry breaking effect is at its maximal. This configuration can maximize the $H_{z,\text{DMI}}^{\text{sat}}$ that switches $\mathbf{M}_{\text{PMA}}$ dictated by $\mathbf{M}_{\text{IMA}}$, which is in turn controlled by the applied current $I_{\text{write}}$ polarity. As shown in Fig. 4(a), current-induced field-free switching can be achieved for both $\mathbf{M}_{\text{PMA}}$ and $\mathbf{M}_{\text{IMA}}$. As further summarized in Fig. 4(b), the critical switching currents $I_c$ share the same trend for both $\mathbf{M}_{\text{IMA}}$ and $\mathbf{M}_{\text{PMA}}$, which increase with higher stack order. This suggests that the moments in both layers can be electrically switched simultaneously due to the coupling of i-DMI. However, as summarized in Fig. 4(c), the current-driven switching percentage of $\mathbf{M}_{PMA}$ drops significantly from 70% to 2% as the stacking number n increased, while that of $\mathbf{M}_{IMA}$ remains fairly constant ~ 100%. The almost constant switching ratio for the in-plane Co layer ($\mathbf{M}_{IMA}$) is ascribed to its type-$y$ SOT switching nature ($I_{write} \perp \mathbf{M}_{IMA}$), which is inherently switchable in a field-free fashion [57-59] and unaffected by increasing the stacking number. In contrast, the fraction of the magnetic domains that can be switched in the PMA [Pt/Co]$_n$ multilayers ($\mathbf{M}_{PMA}$) becomes smaller as increasing n due to the attenuation of i-DMI strength by the weakened coupling between $\mathbf{M}_{IMA}$ and $\mathbf{M}_{PMA}$. This trend is also consistent with our observation that the decreased $H_{z,\text{DMI}}^{\text{sat}}$ cannot fully overcome the coercive field $H_c$ of $\mathbf{M}_{PMA}$ with increasing n (Fig. 2(c)). Note that the SOT that drives $\mathbf{M}_{IMA}$ switching should be mainly originated from the spin Hall current produced in the Pt(2.2) layer. The larger $I_c$ required for samples with higher n is mainly due to additional current shunting with greater overall multilayers thickness, and the value of switching current density $J_{write}$ is similar among all main samples ($J_{write}$ for samples with n = 1, 2, 3, 4 are $2.7 \times 10^{11}$, $3.5 \times 10^{11}$, $3.5 \times 10^{11}$, and $3.6 \times 10^{11}$ A/m$^2$ for PMA switching, and $2.2 \times 10^{11}$, $3.0 \times 10^{11}$, $3.1 \times 10^{11}$, and $3.2 \times 10^{11}$ A/m$^2$ for IMA switching, respectively). The required current density shows no significant variation, suggesting the switching mechanism did not change between different samples (regardless of the $\mathbf{M}_{PMA}$ switching percentage). It is also important to note that the tilted anisotropy plus SOT scenario [14,16] is hardly the main mechanism for the field-free switching observed here, since the switching percentage of $\mathbf{M}_{PMA}$ is the lowest for the sample with the largest tilted angle $\theta_{ani}$ (see Table I).

# VI. Damping-like SOT characterizations

To further shed light on the role of SOT and quantify its contribution in the observed field-free switching of $\mathbf{M}_{\text{PMA}}$, we perform current-induced loop shift measurement to evaluate the SOT-induced effective field $H_z^{\text{eff}}$ in these heterostructures. This is done by applying $I_{\text{DC}}$ along $x$ direction of the Hall bar device to exert a sizable damping-like SOT acting on the Néel domain wall (DW) moments in $\mathbf{M}_{\text{PMA}}$, which translates to an effective field along $z$-direction if a symmetry-breaking $x$-direction in-plane field $H_x$ is also applied and fully aligns the DW moments along $x$ [2,60] (overcoming the classical interfacial DMI effective field). A net $H_z^{\text{eff}}$ proportional to $I_{\text{DC}}$ then can be estimated from the shifted out-of-plane hysteresis loops, as shown in Fig. 5(a) (representative loop shifts for $I_{\text{DC}} = \pm 2.5$ mA) and (b) (extracted $H_z^{\text{eff}}$ vs. $I_{\text{DC}}$) for the n = 1 sample under an in-plane field $H_x = 600$ Oe.

The $H_x$ ($\varphi_H = 0°$) and $H_y$ ($\varphi_H = 90°$) dependence of $H_z^{\text{eff}}/I_{\text{DC}}$ are further summarized in Fig. 5(c), which follow the features of conventional heavy metal (HM)/ferromagnetic metal (FM) bilayer structures [2,16]; $H_z^{\text{eff}}/I_{\text{DC}}$ appears to be an odd function with respect to $H_x$, first increases with magnetic field then saturates. In all samples, $H_x$ of ~ 600 Oe is enough to saturate interfacial DMI field to provide maximized $H_z^{\text{eff}}/I_{\text{DC}}$, and there is no observable $H_z^{\text{eff}}/I_{\text{DC}}$ when $H_y$ is applied. However, even though there is indeed a finite contribution of $H_z^{\text{eff}}$ enforcing on $\mathbf{M}_{\text{PMA}}$ caused by the spin current, the value is far too small to be a deterministic factor to achieve field-free switching with these SOTs assisted by the tilted anisotropy. As shown in Fig. 5(d), even the highest $H_z^{\text{eff}}/I_{\text{DC}}$ is lower than 3 Oe/mA when the interfacial DMI field is fully overcome. By taking the device geometries into account, we obtain $H_z^{\text{eff}}/J_{\text{DC}}$ = 11, 5.8, 5.1, and 2.1 Oe/$10^{11}$A·m$^{-2}$ for sample with n =1, 2, 3, and 4, respectively, where $J_{\text{DC}}$ represents the current density flowing through the Hall bar channels. These values are several times smaller than those reported in other works with Pt-based wedge system [2,16]. Any attempts of $H_x$ assisted current-induced switching measurements also results in non-switching, due to the low $H_z^{\text{eff}}/J_{\text{DC}}$, and the pinned **M**$_{\text{IMA}}$. These results confirm our assumption that **M**$_{\text{PMA}}$ switching mainly relies on the i-DMI and its coupling to the **M**$_{\text{IMA}}$ rather than the tilted anisotropy and the SOT acting upon it.

More recently, several studies also reported on the existence of unconventional spin currents with z-spin polarization ($\sigma_z$) that can be utilized to achieve field-free switching of **M**$_{\text{PMA}}$ in IMA-FM/HM/PMA-FM trilayer structures [19,61]. One possible scenario is that when the classical y-spin $\sigma_y$ from HM scattered and polarized within the IMA-FM, spin-orbit precession (SOP) induced $\sigma_z$ would be generated following the symmetry of $\boldsymbol{\sigma}_y \times \mathbf{m}_x$, providing a finite $H_z^{\text{eff}}$ under zero field condition ($H_x$ = 0 Oe), where **m**$_x$ indicates magnetic moments in IMA-FM. However, $H_z^{\text{eff}}/I_{\text{DC}}$ at $H_x$ = 0 Oe are minimal for all our samples, as shown in Fig. 5(d). $H_z^{\text{eff}}/J_{\text{DC}}$ at $H_x$ = 0 Oe are also extremely low for all samples, as listed in Table 1 with other parameters. This may be ascribed to numerous reasons, including (i) the usage of a strong spin-orbit interaction material Pt(2.2) as the spacer layer, which leads to rapid spin dephasing [62,63]; (ii) insufficient **m**$_x$ in the IMA Co layer since the existing i-DMI will favor **D** (stabilized along $x$) $\perp$ **M**$_{\text{IMA}}$ (stabilized along $y$); or (iii) simply too low a net spin current (limited $\boldsymbol{\sigma}_y$) for the SOP to manifest.

## VII. Conclusions

In summary, we identify two contributions to $H_z^{shift}$ when measuring the out-of-plane hysteresis loops in the presence of a static $H_{IP}$ in a symmetry breaking multilayer system consists of both $\mathbf{M}_{PMA}$ and $\mathbf{M}_{IMA}$: i-DMI coupling and titled magnetic anisotropy. The two mechanisms, however, have drastically different dependences on the [Pt/Co]$_n$ stacking number n. The i-DMI contribution diminishes while the tilted anisotropy gets more significant as increasing n, which suggests that the observed current-induced field-free switching of $\mathbf{M}_{PMA}$ is governed by the i-DMI coupling between $\mathbf{M}_{PMA}$ and $\mathbf{M}_{IMA}$. We further quantify the current-induced SOT effective field $H_z^{eff}$ under various conditions, which exclude the existence of unconventional spins (SOP induced $\sigma_z$) and the possibility of field-free switching with conventional SOT assisted by the tilted anisotropy. Our results therefore point out the importance of precise analysis on (i) asymmetric interlayer exchange coupling such as i-DMI ($\mathbf{D}$ and $H_{z,DMI}^{sat}$), (ii) tilted magnetic anisotropy ($\theta_{ani}$), as well as (iii) SOT contributions ($H_z^{eff}/I_{DC}$) in understanding the field-free switching in magnetic heterostructures with structural symmetry breaking.

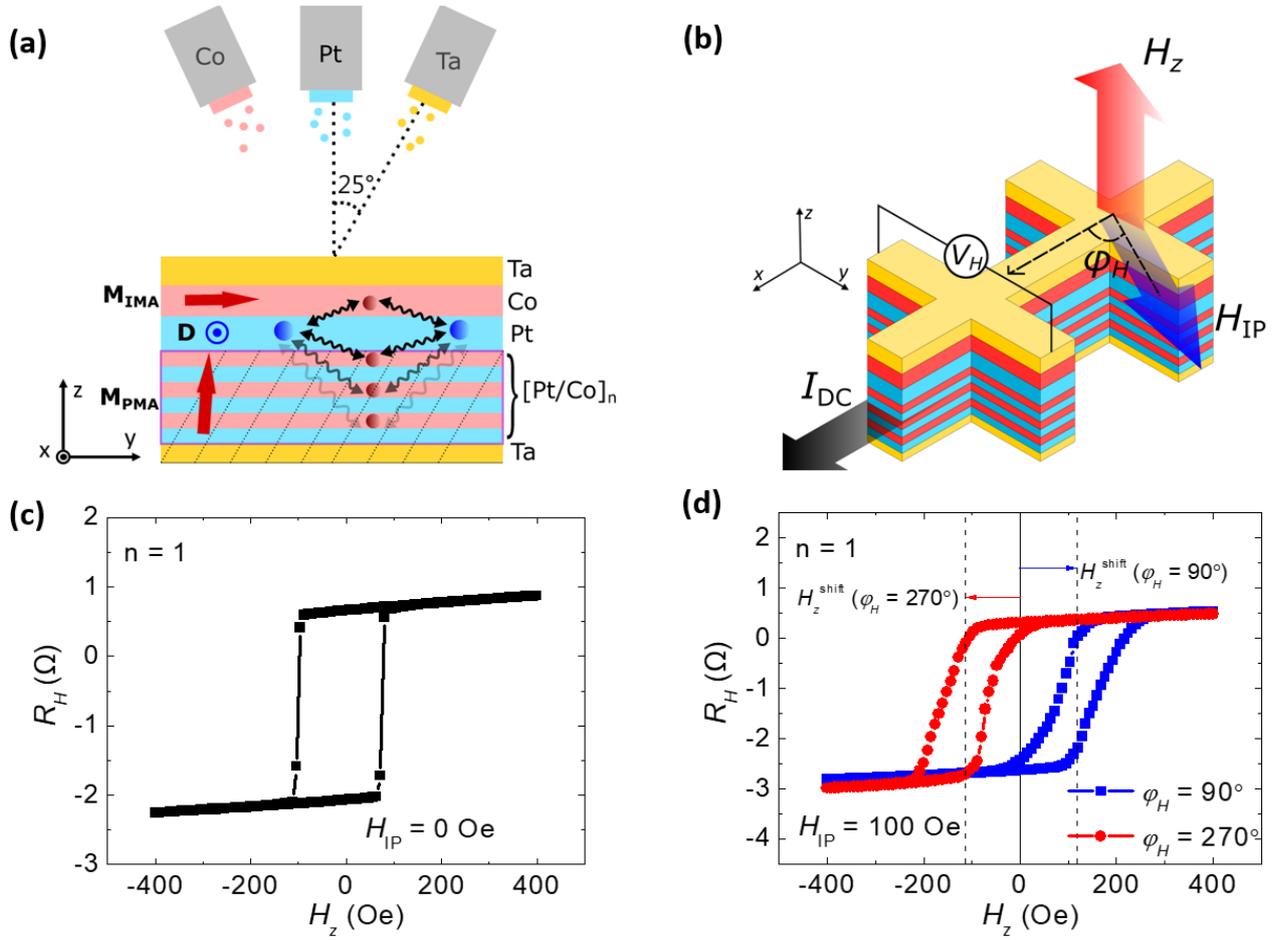

FIG. 1. (a) Schematic illustration of the [Pt/Co]$_n$/Pt/Co multilayer system with broken symmetry, which shows the relative orientations of **M**$_{IMA}$ (in-plane magnetized Co), **M**$_{PMA}$ ([Pt/Co]$_n$ with PMA), and i-DMI vector **D**. The dashed lines represent the obliquely-grown direction of the Ta buffer and the [Pt/Co]$_n$ multilayers. (b) Experimental setup to quantify the effective field induced by i-DMI coupling with a micron-sized Hall bar device. Representative out-of-plane hysteresis loops of **M**$_{PMA}$ in the sample n = 1 with (c) no in-plane field applied ($H_{IP}$= 0 Oe) and (d) $H_{IP}$= 100 Oe for $\varphi_H$ = 90° and 270°. $H_z^{shift}$ represents the magnitude of which the hysteresis loop center has shifted away from $H_z$ = 0.

| n | $H_C$ (Oe) | $H_{z,\text{DMI}}^{\text{sat}}$ (Oe) | $\theta_{\text{ani}}$ (deg) | zero-field switching percentage (%) | zero-field $H_z^{\text{eff}}/J_{\text{DC}}$ (Oe/$10^{11}$ A·m$^{-2}$) |
|---|---|---|---|---|---|
| 1 | 88 | 101 | 2.4 ± 0.13 | 70 | 0.7 ± 1.30 |
| 2 | 115 | 82 | 4.4 ± 0.20 | 40 | 4.3 ± 0.52 |
| 3 | 143 | 61 | 5.5 ± 0.36 | 6 | 2.6 ± 0.67 |
| 4 | 151 | 34 | 5.6 ± 0.50 | 2 | 0.8 ± 1.01 |

Table 1. Summary of stacking number n dependence of the measured and the estimated quantities for $\mathbf{M}_{\text{PMA}}$ ([Pt/Co]$_n$). The uncertainties originate from the standard errors in fittings.

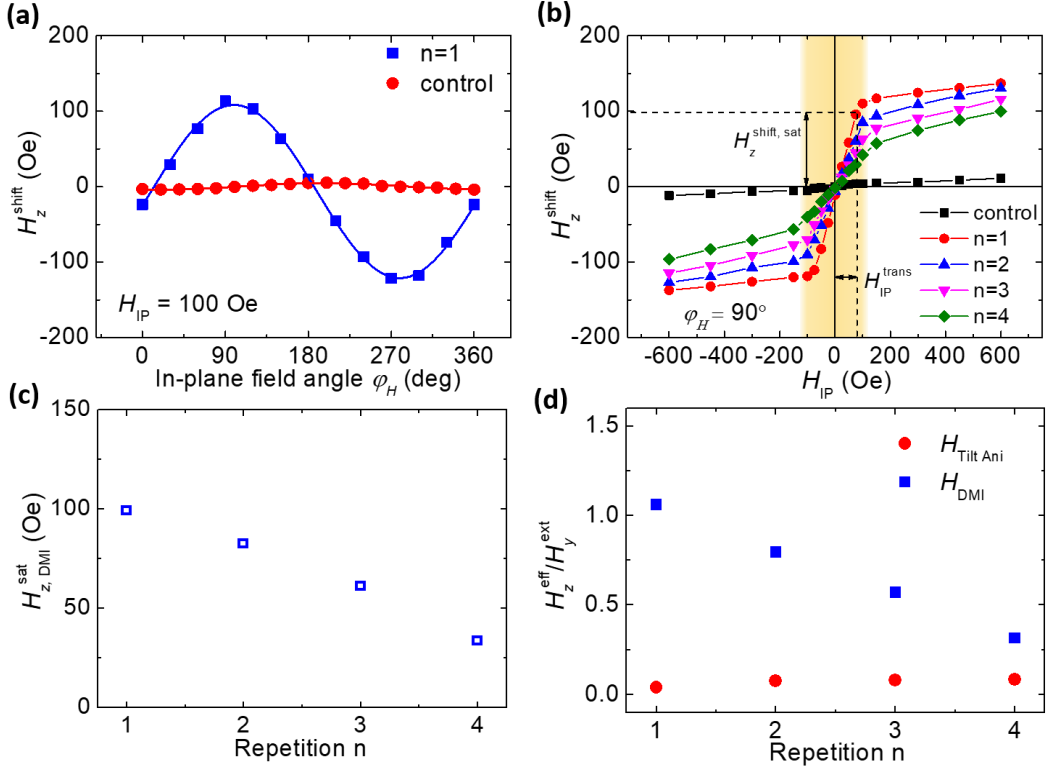

FIG. 2. (a) Measured $H_z^{shift}$ of $\mathbf{M}_{PMA}$ ([Pt/Co]$_n$) as a function of $\varphi_H$ for the Hall bar device with n = 1 (obliquely deposited) and control sample (uniformly deposited with full rotation of the sample holder) under $H_{IP}$ = 100 Oe. (b) $H_z^{shift}$ under different value $H_{IP}$ with $\varphi_H = 90°$ for the control and all obliquely-grown samples. For obliquely-grown samples, $H_z^{shift}$ can be divided into two parts, the i-DMI coupling dominating regime (the shaded section with $H_{IP} < H_{IP}^{trans}$) and the tilted anisotropy dominating regime (the white section with $H_{IP} > H_{IP}^{trans}$). $H_z^{shift, sat}$ and $H_{IP}^{trans}$ for the n = 1 sample are indicated by the black double headed arrows. (c) The i-DMI effective field $H_{z, DMI}^{sat}$ and (d) the extracted values of $H_z^{eff}/H_{IP}$ contributed by i-DMI coupling ($H_{DMI}$) and tilted anisotropy ($H_{Tilt Ani}$) as functions of the stacking repetition number n.

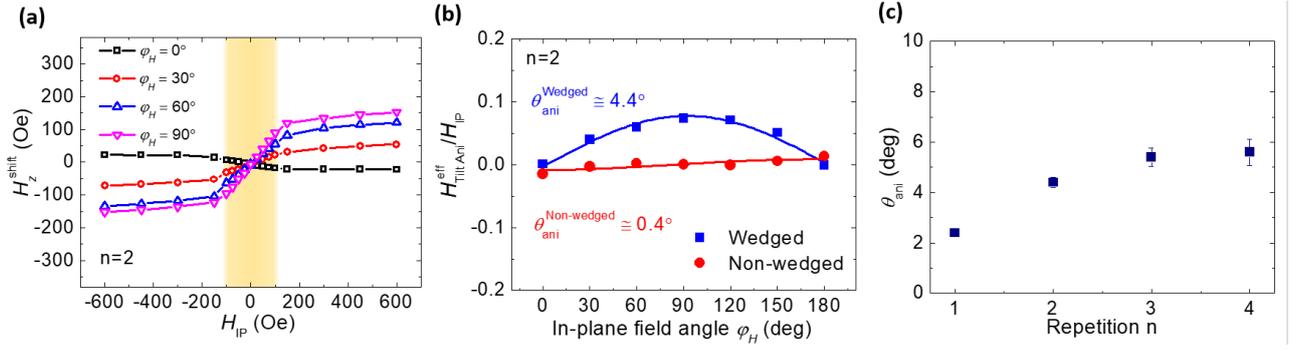

FIG. 3. Estimation of the easy axis tilted angle $\theta_{ani}$ (away from film normal) through loop shift measurements. (a) $H_z^{shift}$ variation under different $\varphi_H$ and $H_{IP}$ for the sample with n = 2. (b) The ratios $H_{Tilt\ Ani}^{eff}/H_{IP}$ extracted from $\frac{H_z^{shift}}{H_{IP}}$ in the white regions (titled anisotropy dominated regime) under different $\varphi_H$ for the n = 2 samples grown obliquely (wedged, blue squares) and uniformly (non-wedged, red circles), from which $\theta_{ani}$ can be extracted using Eq. (1). (c) $\theta_{ani}$ for all obliquely-grown samples.

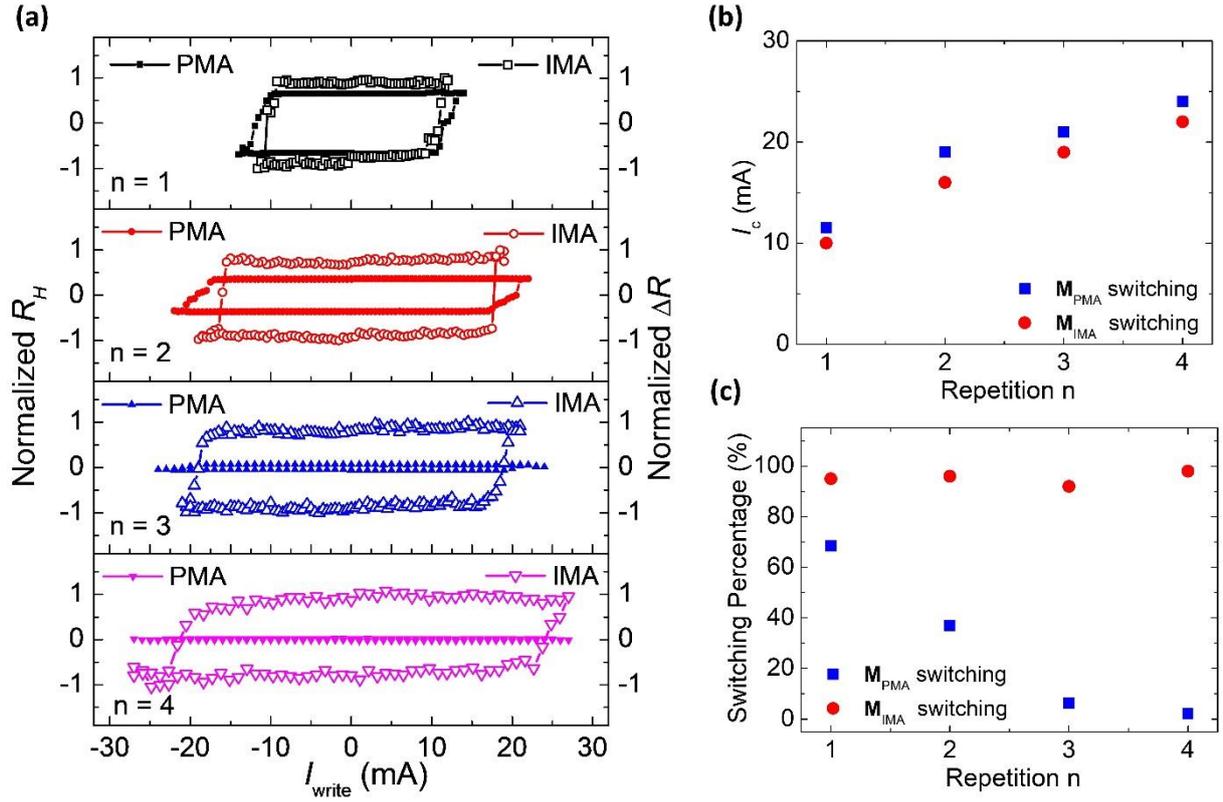

FIG. 4. Field-free switching of $M_{PMA}$ and $M_{IMA}$ coupled via i-DMI. (a) Current-induced magnetization switching loops for Hall bar samples with stacking number n = 1 to 4 (While having a low switching percentage, current switching loop of $M_{PMA}$ for n = 4 could be measured repeatedly). The solid data points represent $M_{PMA}$ measured through $R_H$ and the open data points represent $M_{IMA}$ sensed by the UMR ($\Delta R$). Stacking number dependence of (b) critical switching current $I_c$ and (c) switching percentage for both $M_{PMA}$ and $M_{IMA}$ of the four samples.

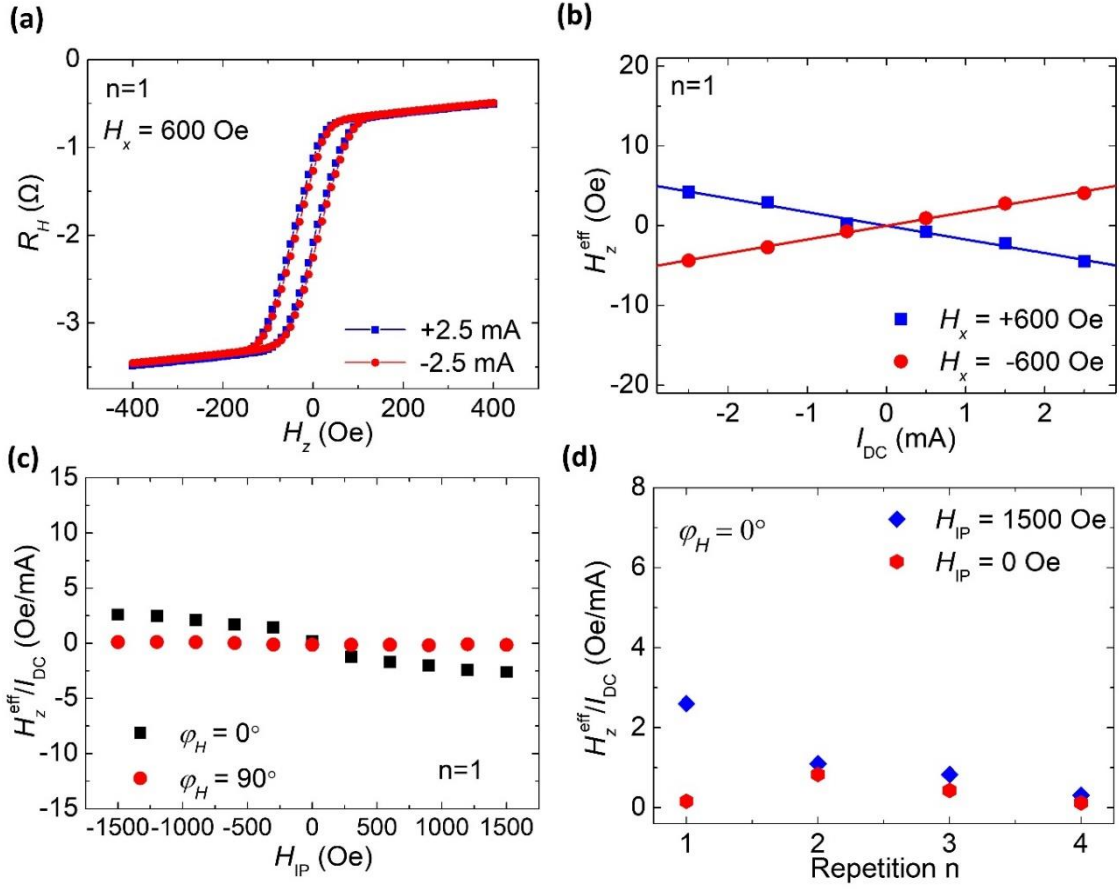

FIG 5. SOT characterization results measured by current-induced hysteresis loop shift measurements. (a) Out-of-plane hysteresis loops of $\mathbf{M}_{PMA}$ sensed by $R_H$ for a Hall bar sample with n = 1 under $I_{DC} = \pm 2.5$ mA and $H_x = 600$ Oe. (b) The SOT-induced $H_z^{eff}$ versus $I_{DC}$ for sample n = 1 under $H_x = \pm 600$ Oe. (c) $H_z^{eff}/I_{DC}$ versus $H_{IP}$ with $\varphi_H = 0°$ and $90°$ ($H_x$ and $H_y$) for sample n = 1. (d) $H_z^{eff}/I_{DC}$ vs. the stacking repetition number n under $H_{IP} = 0$ and 1500 Oe with $\varphi_H = 0°$.

**Appendix A: Relations between the Pt spacer thickness and i-DMI effective field**

In order to find an optimal Pt spacer layer thickness to serve as the basis for analyses and to investigate the i-DMI spacer thickness dependence, we have separately fabricated a series of samples which all have a single Pt(1)/Co(0.8) stack as the PMA layer, and the Pt spacer thickness stacked on top ranged from 1.8 nm to 2.5 nm (similar to the n = 1 sample in the main text). The i-DMI effective fields of the samples with different Pt spacer thicknesses are quantified using the identical field-sweep protocol described in the main text (as seen in Fig. 1(b)). Fig. A1 (a) shows the $H_z^{\text{shift}}$ as a function of $\varphi_H$, and Fig. A1 (b) compiles both the **D** vector's angle, $\varphi_D$, and $H_{z,\text{DMI}}^{\text{sat}}$ as functions of Pt space thickness. The **D** vector's direction shows great agreement with the representative data of Fig. 2(a) with $\varphi_D$ universally close to 0°. For Pt thickness ≥ 2.0 nm, $H_{z,\text{DMI}}^{\text{sat}}$ shows a quasi-monotonic decay with increasing the Pt thickness, that can be fitted by (Pt thickness)$^{-1}$, in agreement with the nature of i-DMI being an interfacial effect. However, $H_{z,\text{DMI}}^{\text{sat}}$ decreases drastically when Pt thickness is < 2.0 nm. Previous observations by Avci. *et al* [48] have shown a very similar trend with regard to the spacer thickness, and it was argued that the spacer thickness dependence deviates from a damped oscillation due to the total DMI being averaged out by all three-site interactions between PMA and IMA magnetizations. This deterioration at low Pt thicknesses could also be related to the inferior interfacial quality, such as the Pt layer becoming discontinuous islands rather than a continuous film, thus weakening the Pt/Co interface's spin orbit coupling and subsequently the i-DMI strength [64].

Aside from a sizable $H_{z,\text{DMI}}^{\text{sat}}$, robust PMA characteristics are also required for current-induced hysteresis loop shift and current-induced magnetization switching measurements. $H_c$ and the ratio of out-of-plane remnant/saturated magnetization ($M_R/M_S$) as functions of Pt thickness are shown in Fig. A1(c), with rapid deterioration of PMA evident when Pt thickness is less than 2.0 nm, which may also be related to the decreased interfacial spin orbit coupling [65]. Judging from the results in Fig. A1(b) and(c), we concluded that Pt thickness being 2.2 nm strikes a good balance between a sizable $H_{z,\text{DMI}}^{\text{sat}}$ and robust PMA characteristics.

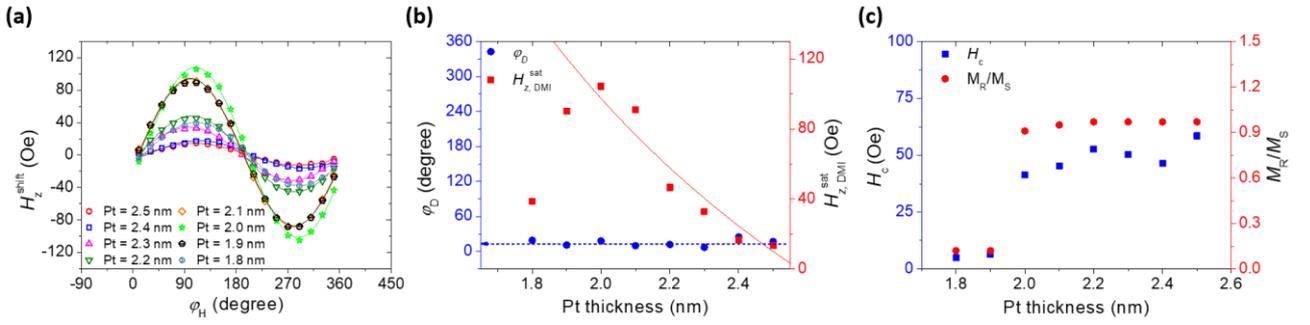

FIG. A1. (a) $H_z^{\text{shift}}$ at different Pt spacer thicknesses with varying $\varphi_H$, solid lines are sine fits to the data. (b) Compilations of both $\varphi_D$ and $H_{z,\text{DMI}}^{\text{sat}}$ as functions of Pt thicknesses, extracted from sine fits in (a). Red solid line denotes a (Pt thickness)$^{-1}$ fitting to $H_{z,\text{DMI}}^{\text{sat}}$, between Pt thickness = 2.0 nm to 2.5 nm. (c) Characterization of perpendicular direction (out-of-plane) coercivity and out-of-plane remnant/saturated magnetization as functions of Pt spacer thickness.

## Appendix B: i-DMI induced AMR loop shift of the IMA layer

From the i-DMI Hamiltonian described in the main text, it is suggested that a reciprocal i-DMI effective field originated from $M_{PMA}$ is also exerted on $M_{IMA}$ along $\pm y$ direction. To prove the existence of such effective field, we obtained the $M_{IMA}$ response under the influence of i-DMI by measuring the anisotropic magnetoresistance (AMR) response of the IMA Co layer. Since the i-DMI effective field exerted on $M_{IMA}$ takes the form of $\mathbf{H}_{DMI} = \mathbf{D} \times \mathbf{M}_{PMA}$, the AMR loops should show measurable shifts in the $\pm y$ direction (with $\mathbf{D}$ lies approximately toward $0°$ and $M_{PMA}$ fixed toward $\pm z$). Fig. A2 (a) and (b) show the AMR shifts of n = 1 and n = 4 samples, with the measured i-DMI effective field exerted on $M_{IMA}$ to be 120 Oe and 65 Oe, respectively. The decreasing trend agrees with the gradually decreasing $H_{z,DMI}^{sat}$ when n is increased from 1 to 4, as reported in the main text for $M_{PMA}$. Note that due to the sizable $H_{IP}$ used to properly align $M_{IMA}$, the i-DMI effective field enacting on $M_{IMA}$ does not modify the results in the main text.

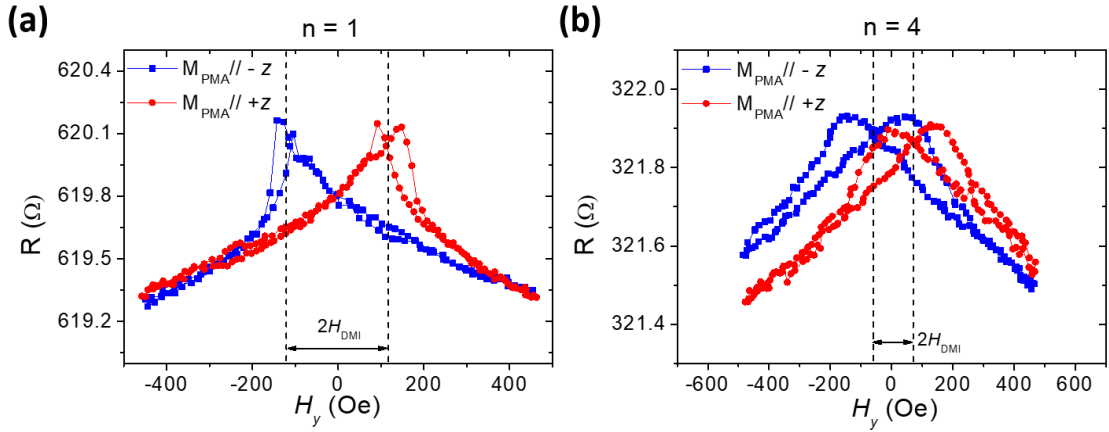

FIG. A2. Representative shifted AMR loops due to i-DMI effective field acting on the $M_{IMA}$ (in-plane Co layer) for (a) n = 1 sample, and (b) n = 4 sample. The $H_{DMI}$ acting on either samples is obtained by fixing $\mathbf{M}_{PMA}$ toward $\pm z$ when measuring the AMR loops.